\documentstyle[psfig]{mn}
\newif\ifAMStwofonts
\def\simless{\mathbin{\lower 3pt\hbox
     {$\rlap{\raise 5pt\hbox{$\char'074$}}\mathchar"7218$}}}   %< or of order
\def\simmore{\mathbin{\lower 3pt\hbox
     {$\rlap{\raise 5pt\hbox{$\char'076$}}\mathchar"7218$}}}   %> or of order
\ifoldfss
  \ifCUPmtlplainloaded \else
    \NewTextAlphabet{textbfit} {cmbxti10} {}
    \NewTextAlphabet{textbfss} {cmssbx10} {}
    \NewMathAlphabet{mathbfit} {cmbxti10} {} % for math mode
    \NewMathAlphabet{mathbfss} {cmssbx10} {} %  "   "    "
  \fi
  \ifAMStwofonts
    \ifCUPmtlplainloaded \else
      \NewSymbolFont{upmath} {eurm10}
      \NewSymbolFont{AMSa} {msam10}
      \NewMathSymbol{\upi}     {0}{upmath}{19}
      \NewMathSymbol{\umu}     {0}{upmath}{16}
      \NewMathSymbol{\upartial}{0}{upmath}{40}
      \NewMathSymbol{\leqslant}{3}{AMSa}{36}
      \NewMathSymbol{\geqslant}{3}{AMSa}{3E}

      \let\leq=\leqslant 
       
    \fi
  \fi
\fi % End of OFSS

\ifnfssone
  \newmathalphabet{\mathit}
  \addtoversion{normal}{\mathit}{cmr}{m}{it}
  \addtoversion{bold}{\mathit}{cmr}{bx}{it}
  \newmathalphabet{\mathbfit} % math mode version of \textbfit{..}
  \addtoversion{normal}{\mathbfit}{cmr}{bx}{it}
  \addtoversion{bold}{\mathbfit}{cmr}{bx}{it}
  \newmathalphabet{\mathbfss} % math mode version of \textbfss{..}
  \addtoversion{normal}{\mathbfss}{cmss}{bx}{n}
  \addtoversion{bold}{\mathbfss}{cmss}{bx}{n}
  \ifAMStwofonts
    \ifCUPmtlplainloaded \else
      \UseAMStwoboldmath
      \makeatletter
      \new@mathgroup\upmath@group
      \define@mathgroup\mv@normal\upmath@group{eur}{m}{n}
      \define@mathgroup\mv@bold\upmath@group{eur}{b}{n}
      \edef\UPM{\hexnumber\upmath@group}
      \new@mathgroup\amsa@group
      \define@mathgroup\mv@normal\amsa@group{msa}{m}{n}
      \define@mathgroup\mv@bold\amsa@group{msa}{m}{n}
      \edef\AMSa{\hexnumber\amsa@group}
      \makeatother
      \mathchardef\upi="0\UPM19
      \mathchardef\umu="0\UPM16
      \mathchardef\upartial="0\UPM40
      \mathchardef\leqslant="3\AMSa36
      \mathchardef\geqslant="3\AMSa3E

      \let\leq=\leqslant 

    \fi
  \fi
\fi % End of NFSS release 1

\ifnfsstwo
  \DeclareMathAlphabet{\mathbfit}{OT1}{cmr}{bx}{it}
  \SetMathAlphabet\mathbfit{bold}{OT1}{cmr}{bx}{it}
  \DeclareMathAlphabet{\mathbfss}{OT1}{cmss}{bx}{n}
  \SetMathAlphabet\mathbfss{bold}{OT1}{cmss}{bx}{n}
  \ifAMStwofonts
    \ifCUPmtlplainloaded \else
      \DeclareSymbolFont{UPM}{U}{eur}{m}{n}
      \SetSymbolFont{UPM}{bold}{U}{eur}{b}{n}
      \DeclareSymbolFont{AMSa}{U}{msa}{m}{n}
      \DeclareMathSymbol{\upi}{0}{UPM}{"19}
      \DeclareMathSymbol{\umu}{0}{UPM}{"16}
      \DeclareMathSymbol{\upartial}{0}{UPM}{"40}
      \DeclareMathSymbol{\leqslant}{3}{AMSa}{"36}
      \DeclareMathSymbol{\geqslant}{3}{AMSa}{"3E}

      \let\leq=\leqslant 

    \fi
  \fi
\fi % End of NFSS release 2

\ifCUPmtlplainloaded \else
  \ifAMStwofonts \else % If no AMS fonts
    \def\upi{\pi}
    \def\umu{\mu}
    \def\upartial{\partial}
  \fi
\fi

\title[kHz QPOs in Sco X-1]
{The Harmonic and Sideband Structure of the Kilohertz Quasi-Periodic
Oscillations in Sco X-1}
\author[M. M\'endez and M. van der Klis]
       {Mariano M\'endez$^{1,2}$ and Michiel van der Klis$^{1}$\\
        $^1$ Astronomical Institute `Anton Pannekoek',
        University of Amsterdam and Center for High-Energy Astrophysics,
        Kruislaan 403,\\
        NL-1098 SJ Amsterdam, The Netherlands.\\
        $^2$ Facultad de Ciencias Astron\'omicas y Geof\'{\i}sicas,
        Universidad Nacional de La Plata, Paseo del Bosque S/N,\\
        1900 La Plata, Argentina.}
\date{Accepted 2000 June .
      Received 2000 February 28;
      in original form 2000 February 28}

\pagerange{\pageref{firstpage}--\pageref{lastpage}}
\pubyear{2000}

\begin{document}

\maketitle

\label{firstpage}

\begin{abstract}
We use data from the {\em Rossi X-ray Timing Explorer} to search for
harmonics and sidebands of the two simultaneous kilohertz
quasi-periodic oscillations (kHz QPOs) in Sco X-1. We do not detect any
of these harmonics or sidebands, with 95\,\% confidence upper limits to
their power between $\sim 1$\,\% and $\sim 10$\,\% of the power of the
upper kHz QPO. The oscillations produced at these frequencies may be
attenuated in a scattering corona around the neutron star. We find that
upper limits to the unattenuated power of some of the strongest
theoretically predicted harmonics and sidebands are as low as $\sim 2$\,\% of the unattenuated power of the high-frequency QPO in Sco X-1.
\end{abstract}

\begin{keywords}
accretion, accretion discs
-- stars: neutron
-- stars: Sco X-1
-- X-rays: stars
\end{keywords}

\section{Introduction}
\label{intro}

It is four years now since the kilohertz quasi-periodic oscillations
(kHz QPOs) were discovered in the persistent flux of Scorpius X-1
\cite{vdk_iauc} and 4U\,1728--34 \cite{stroh_iauc}. In the meantime,
similar kHz QPOs have been seen in some 20 other low-mass X-ray
binaries (LMXBs; see van der Klis 2000 for a review). These QPOs often
appear in pairs, with frequencies $\nu_{1}$ and $\nu_{2}$ ($\nu_{2} >
\nu_{1}$) between $\sim 400$ and $\sim 1300$ Hz, which in a given
source can shift by a few hundred Hz, apparently as a function of mass
accretion rate.

Most of the models proposed so far assume that one of the kHz QPOs
reflects the Keplerian orbital motion at some preferred radius in the
accretion disc (e.g., Miller, Lamb \& Psaltis 1998; Stella \& Vietri
1999; Osherovich \& Titarchuk 1999), but there are other explanations
as well (Klein et al. 1996a,b; Jernigan, Klein \& Arons 2000). In
recent discussions (Lamb \& Miller 1999; Stella 1999; Psaltis 1999) it
was emphasized that empirical discrimination between two of the leading
classes of models is possible in principle by studying the harmonic and
sideband structure of the kHz QPOs. Here we concentrate only on these
two model classes

In the `sonic-point' model (SPM; Miller et al. 1998), the QPO at
$\nu_{2}$ (the upper QPO) is produced at the radius where the radial
flow velocity in the disc turns from subsonic to supersonic (the sonic
radius), and the QPO at $\nu_{1}$ (the lower QPO) originates by a beat
between the upper QPO and the spin frequency of the neutron star. In
the `relativistic-precession' model (RPM; Stella \& Vietri 1999) the
QPO at $\nu_{2}$ is also assumed to be Keplerian, but the QPO at
$\nu_{1}$ is produced by the apsidal precession of a slightly
non-circular inner accretion disc. The QPO frequencies in the RPM are
calculated for test particles in purely geodesic relativistic motion,
i.e., neglecting the hydrodynamical and radiative effects of the
accretion flow. However, Psaltis \& Norman \shortcite{psaltis_norman}
have recently proposed a dynamical model in which the QPOs are produced
by oscillations in the accretion disk. In this model, which we will
call `transition-radius' model (TRM), there is a transition radius in
the accretion disc that acts as a band-pass filter with resonances near
the orbital and periastron-precession frequencies.

Besides the main peaks at $\nu_{1}$ and $\nu_{2}$, the SPM and the TRM
predict other (weaker) harmonics and sidebands of these QPOs, at
specific frequencies. For instance, the SPM predicts a relatively
strong harmonic of the lower QPO at $2 \nu_1$ (see Table 3 of Miller et
al. 1998 for a list of other sideband peaks predicted by the SPM),
whereas the TRM predicts a sideband at $2 \nu_{2} - \nu_{1}$ (cf.
eq.~[29] in Psaltis \& Norman 2000). In principle, the detection of a
QPO at $2 \nu_1$ and a non-detection of a QPO at $2 \nu_{2} - \nu_{1}$
would tend to rule out the TRM, whereas the detection of a QPO at $2
\nu_{2} - \nu_{1}$ and a non-detection of a QPO at $2 \nu_1$ would tend
to rule out the SPM \cite{miller_bologna}.

In this Letter, we use data from the {\em Rossi X-ray Timing Explorer
(RXTE)} to search for the predicted harmonics and sidebands in the
power spectrum of the kHz QPO source Sco X-1. Because of its very high
flux, Sco X-1 has extremely significant kHz QPOs, and hence a very
sensitive study of any harmonic structure is possible. We do not detect
any of these secondary QPOs. The fact that we detect the kHz QPOs at
$\nu_{1}$ and $\nu_{2}$, but none of these other peaks sets severe
constraints on the models currently proposed to explain the kHz QPOs in
LMXBs.

\section[]{Observations}
\label{observations}

We used data from the Proportional Counter Array (PCA; Jahoda et al.
1996) on board {\em RXTE} (Bradt, Rothschild \& Swank 1993) taken on
1996 February 14, 18, 19, 1996 May 24 -- 28, 1997 March 15, 1997 April
18 -- 24, 1997 August 22, 1998 January 2 -- 8, 1998 February 27, 28,
1998 May 30, 31, 1998 June 1, 2, 1998 July 2 -- 5, 1999 January 6, 8 --
11 and 13 -- 16. Each observation (i.e., each part of the data with a
unique {\em RXTE} ID number) consists of data blocks of $\sim 60$ s to
$\sim 3,700$ s interrupted by passages of the satellite through the
South Atlantic Anomaly and occultations of the source by the Earth. The
total usable time was $\sim 630$ ks.

To avoid detector safety triggers, telemetry saturation, and to reduce
the dead-time effects produced by the high count rate of Sco X-1, some
observations were carried out with the source slightly off-axis, with
some of the five proportional counter units of the PCA switched off,
recording only photons detected by the upper anode chain of the PCA,
recording only photons from a limited energy range, or using a
combination of these constraints. In all cases, high-time resolution
data were available with a time resolution of 0.25 ms or better. For
our analysis below we combined all the available data, irrespective of
whether they were collected using any of the above observational
constraints. When single and double-event data were recorded in
parallel (see van der Klis et al. 1996b), we combined them off-line to
enhance the sensitivity.

\section{Analysis and Results}
\label{analysis}

We divided the high time resolution data into 16 s segments, and
produced a power spectrum for each of these segments up to a Nyquist
frequency of 2048 Hz (for $\sim 50$\,\% of the data we also produced
power spectra up to a Nyquist frequency of 4096 Hz). For 69\,\% of the
observations the power spectra were calculated using the full PCA
energy band. For the rest of the power spectra we used data from
selected energy bands (24\,\% of the power spectra were calculated
between $5-18$ keV, 5\,\% between $5-60$ keV, and 2\,\% between $2-18$
keV) because we found that the kHz QPOs were more significant in those
energy bands, or because those were the only energy bands available.
Finally, we averaged together groups of 8 contiguous 16 s power spectra
to produce average power spectra.

We searched these average power spectra for kHz QPOs, at frequencies
$\simmore 250$ Hz. We did this by first identifying those power spectra
that showed a strong QPO, and visually estimating its frequency,
$\nu_{0}$. When two QPOs were present in the power spectra, we always
picked the one at higher frequency. It turned out that in the cases
where only one kHz QPO was visible, it was always the upper kHz QPO
(see below). We then fitted the power spectra, in the range $\nu_{0} -
100$ Hz to $\nu_{0} + 100$ Hz, using a function consisting of a
constant, a power law and one Lorentzian. We discarded the power
spectra for which the QPOs were less than $3 \sigma$ significant. We
note that in this manner we might have discarded data with weak QPOs
that could have been detected averaging together more data. There were
1,384 average power spectra with significant kHz QPOs, equivalent to
177,152 s of data.

\begin{figure}
\centerline{\psfig{file=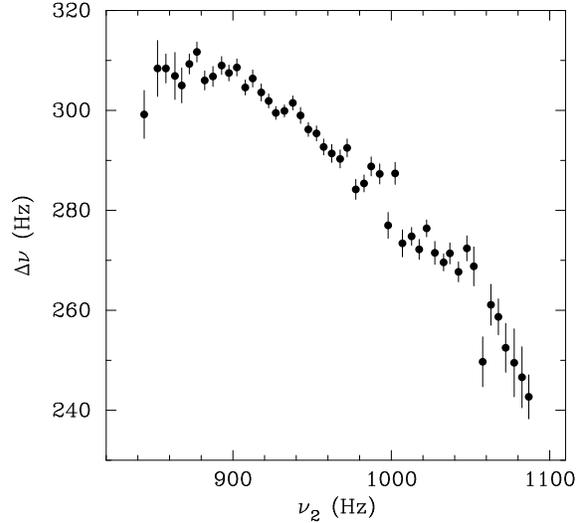,angle=-90,width=7.5cm}}
\caption[fig1.ps]{The frequency separation between the kHz QPOs
in Sco X-1 as a function of the upper kHz QPO frequency.
\label{figdif}
}
\end{figure}

For those observations where we detected only one QPO in the average
power spectra we applied the `shift-and-add' technique
\cite{mendez1608_1} to try and detect the second QPO: Based on the
frequency of the detected QPO, we aligned and averaged all the power
spectra of a single observation. In those cases, this procedure
revealed a second kHz QPO, which in all cases was at a lower frequency
than the QPO peak that we used to align the power spectra, showing that
all the initial frequency measurements in the 1,384 average power
spectra corresponded to the upper kHz QPO.

Next, we measured the frequency separation between the kHz QPOs as a
function of the upper QPO frequency, $\nu_{2}$. We proceeded as
follows: We aligned the 1,384 power spectra using the upper kHz QPO as
a reference; we grouped the data in 49 sets, each of them containing
$\sim 40$ to $\sim 210$ power spectra, such that $\nu_{2}$ did not vary
by more than $5-10$ Hz within each set, and we combined these aligned
spectra to produce an average power spectrum for each set. We fitted
these 49 power spectra in the range $400 - 1300$ Hz using a function
consisting of a constant, a power law and two Lorentzians. The fits
were good, with reduced $\chi^{2} \leq 1.1$ for 279 degrees of freedom,
and the significance of both peaks was always $> 3\sigma$.

Fig. \ref{figdif} shows that $\Delta \nu= \nu_{2} - \nu_{1}$, the
frequency difference between the two QPOs, decreases from $\sim 310$ Hz
to $\sim 240$ Hz, as $\nu_{2}$ increases from $\sim 840$ Hz to $\sim
1100$ Hz. This figure can be compared to fig. 3a of van der Klis et al.
\shortcite{vdk_sco_2}, who first reported the decrease in $\Delta \nu$
with $\nu_{2}$ in Sco X-1. Because we included more data, and because
we used the shift-and-add technique to measure $\Delta \nu$, the errors
are smaller in the present figure than in the previous one, and some
structure, particularly between 900 and 1020 Hz, becomes apparent
(perhaps this structure is related to the `bump' seen in the plot of
$\Delta \nu$ vs. $\nu_{1}$ of 4U\,1608--52 at $\nu_{1} \sim 700$ Hz;
see fig. 3 in M\'endez et al. 1998b).

From Fig. \ref{figdif} we can read off $\nu_{1}$ as a function of
$\nu_{2}$; because we already know $\nu_{2}$ for each average power
spectrum, we can calculate the expected frequencies of hypothetical
signals at multiples of $\nu_{1}$ and $\nu_{2}$, or at frequencies that
are combinations of these two frequencies, for each power spectrum. As
we described in \S \ref{intro}, some of these frequencies are important
in the context of the models proposed to explain the kHz QPOs.

\begin{figure}
\centerline{\psfig{file=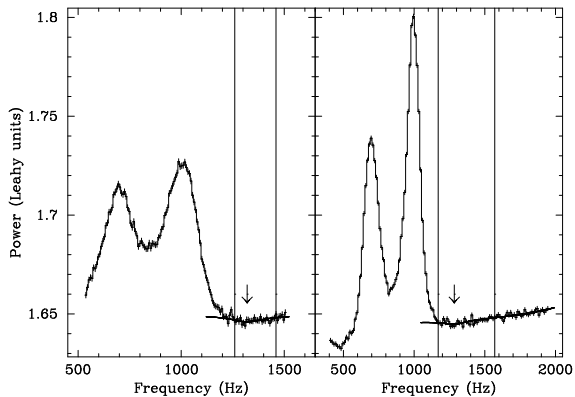,angle=-90,width=7.5cm}}
\caption[fig1.ps]{Shifted-and-averaged power spectra of Sco X-1. The
average power spectrum shown in the left panel was computed by first
aligning the individual power spectra such that a QPO peak at $2
\nu_{1}$ would always end up at 1320 Hz, indicated by the arrow. In the
right panel a QPO peak at $2 \nu_{2} - \nu_{1}$ would similarly end up
at 1284 Hz (arrow). In both panels, the vertical lines show the
frequency range used to fit the data (see text). Notice that the
frequency alignment also altered the apparent shapes and frequencies of
the strong QPO peaks.
\label{pow_spec}
}
\end{figure}

We calculated the following quantities: $2 \nu_{1}$, $2 \nu_{2}$,
$\nu_{2} - \nu_{1}$, $\nu_{2} + \nu_{1}$, $2 \nu_{1} - \nu_{2}$, $2
\nu_{2} - \nu_{1}$, $2 \nu_{1} + \nu_{2}$, $2 \nu_{2} + \nu_{1}$ and $2
(\nu_{2} - \nu_{1})$. We then aligned, in turn, the power spectra on
each of the above frequencies, and averaged them together to try and
detect a signal present at any of these frequencies. We fitted these
shifted and averaged power spectra in a segment of $\sim 400$ Hz that
included the expected frequency, using a constant plus a Lorentzian
with the centroid fixed in each case at the expected frequency, and
with a fixed FWHM of 200 Hz, approximately equal to the sum of the FWHM
of each of the QPOs \cite{vdk_sco_1}. This value is about two times
larger than the largest FWHM measured in the QPOs of this and other
sources, and yields conservative upper limits to the unobserved
harmonics and sidebands; tighter upper limits could be obtained by
fixing the FWHM to a smaller value. No detections result from choosing
a smaller FWHM in the current data set. In a few cases we added a power
law to the fit to take account of a sloping continuum. The shape of the
power spectrum at high frequencies is dominated by dead-time effects
\cite{zhang_dead-time} which, in the case of Sco X-1 at a count rate
that exceeds $\sim 25,000$ c\, s$^{-1}$ \,PCU$^{-1}$, are large and 
not yet sufficiently well understood to predict the shape of the
high-frequency part of the power spectrum accurately.  We note that a
better knowledge of this effect could yield tighter upper limits than
those that we report in this paper.

We did not detect a significant QPO at any of the above frequencies. As
an example, in Fig. \ref{pow_spec} we show two power spectra that were
shifted to $2 \nu_{1}$ and to $2 \nu_{2} - \nu_{1}$, the frequencies of
the main high-frequency peaks (besides the kHz QPOs) predicted by the
models of Miller et al. \shortcite{miller_sonic} and Psaltis \& Norman
\shortcite{psaltis_norman}, respectively.

\begin{table}
\caption{Upper limits to the observed amplitudes of the harmonics and sidebands of the kHz QPOs in Sco X-1
\label{table_UL_obs}
}
\begin{tabular}{ccc}
Frequency                 &
Predicted frequency range &
Upper limit$^{a}$         \\
$2 \nu_{1}$               & 1090 -- 1690 Hz & 0.12 \\
$2 \nu_{2}$               & 1690 -- 2170 Hz & 0.13 \\
$\nu_{2} - \nu_{1}$       &  240 --  300 Hz & 0.30 \\
$\nu_{2} + \nu_{1}$       & 1390 -- 1930 Hz & 0.08 \\
$2 \nu_{1} - \nu_{2}$     &  240 --  600 Hz & 0.26 \\
$2 \nu_{2} - \nu_{1}$     & 1140 -- 1330 Hz & 0.12 \\
$2 \nu_{1} + \nu_{2}$     & 1930 -- 2770 Hz & 0.10 \\
$2 \nu_{2} + \nu_{1}$     & 2230 -- 3020 Hz & 0.10 \\
$2 (\nu_{2} - \nu_{1})$   &  480 --  600 Hz & 0.21 \\
\end{tabular}

\smallskip
$^{a}$ 95\,\% confidence upper limits to the rms fractional amplitude of the QPO at the indicated frequency, in units of the observed rms fractional amplitude of the QPO at $\nu_{2}$. This QPO has an amplitude between 0.6 \% and 2.5 \% \cite{vdk_sco_2}.
\end{table}

Upper limits are shown in Table \ref{table_UL_obs}. To calculate these
upper limits we varied the amplitude of the Lorentzian until the
$\chi^2$ of the fit increased by 2.71 with respect to the best-fitting
value (95\,\% confidence level for a single parameter); the upper
limits we quote in Table \ref{table_UL_obs} represent the change in the
amplitude for this change in the $\chi^2$. Because of the observational
constraints described in \S \ref{observations}, and because we combined
power spectra computed from data in different energy bands, we cannot
give upper limits in units of fractional rms amplitude
\cite{vanderklis95}; instead, all the upper limits in Table
\ref{table_UL_obs} are in units of the observed fractional rms
amplitude of the QPO at $\nu_{2}$. Any peaks at the predicted
frequencies are between 11 and 156 times weaker than the peak at
$\nu_{2}$, which translates into rms amplitude ratios between 0.08 and
0.30.

\section{Discussion}
\label{discussion}

\begin{table*}[H]
\caption{Upper limits to the unattenuated amplitude of the harmonics and sidebands of the kHz QPOs in Sco X-1
\label{table_UL_calc}
}
\begin{tabular}{ccccccc}
Frequency         &
$\tau^{a}$        &
~~$R$ (km)$^{a}$  &
Upper limit$^{b}$ &
$\tau^{a}$        &
~~$R$ (km)$^{a}$  &
Upper limit$^{c}$ \\
%$2 \nu_{1}$             & 0.06 & 0.17 \\
%$2 \nu_{2}$             & 0.10 & 0.28 \\
%$\nu_{2} - \nu_{1}$     & 0.05 & 0.16 \\
%$\nu_{2} + \nu_{1}$     & 0.05 & 0.14 \\
%$2 \nu_{1} - \nu_{2}$   & 0.05 & 0.15 \\
%$2 \nu_{2} - \nu_{1}$   & 0.05 & 0.16 \\
%$2 \nu_{1} + \nu_{2}$   & 0.10 & 0.28 \\
%$2 \nu_{2} + \nu_{1}$   & 0.12 & 0.34 \\
%$2 (\nu_{2} - \nu_{1})$ & 0.05 & 0.15 \\
$2 \nu_{1}$             & $<1$ & ~~336 & 0.15 &
                          8.7  & ~~129 & 0.30 \\
$2 \nu_{2}$             & 6.5  & ~~202 & 0.27 &
                          9.8  & ~~109 & 0.93 \\
$\nu_{2} - \nu_{1}$     & $<1$ & $<10$ & 0.30 &
                          $<1$ & $<10$ & 0.30 \\
$\nu_{2} + \nu_{1}$     & 5.7  & ~~262 & 0.11 &
                          9.1  & ~~121 & 0.35 \\
$2 \nu_{1} - \nu_{2}$   & $<1$ & $<10$ & 0.26 &
                          $<1$ & $<10$ & 0.26 \\
$2 \nu_{2} - \nu_{1}$   & $<1$ & ~~352 & 0.14 &
                          8.3  & ~~138 & 0.23 \\
$2 \nu_{1} + \nu_{2}$   & 7.3  & ~~167 & 0.37 &
                          9.8  & ~~109 & 1.43 \\
$2 \nu_{2} + \nu_{1}$   & 7.8  & ~~151 & 0.54 &
                         10.2  & ~~103 & 2.16 \\
$2 (\nu_{2} - \nu_{1})$ & $<1$ & $<10$ & 0.21 &
                          $<1$ & $<10$ & 0.21 \\
\end{tabular}

\smallskip
$^{a}$ Optical depth and radius of a possible scattering cloud around Sco X-1, obtained by minimizing eq. [\ref{alpha}] at each frequency (see text). These are the values that were used to calculate the corresponding upper limits shown in this Table.\\
$^{b}$ Upper limits for the unattenuated amplitude of the QPO at the indicated frequency for a luminosity oscillation (see eq. [\ref{luminosity}]), in units of the unattenuated amplitude of the QPO
at $\nu_{2}$.\\
$^{c}$ Upper limits for the unattenuated amplitude of the QPO at the indicated frequency for a beaming oscillation (see eq. [\ref{beaming}]), in units of the unattenuated amplitude of the QPO at $\nu_{2}$.\\
In both cases the QPO at $\nu_{2}$ is assumed to be a beaming oscillation (Miller et al. 1998)
\end{table*}

We have measured the frequencies of the two simultaneous kHz QPOs,
$\nu_{1}$ and $\nu_{2}$, in Sco X-1 in a large data set. We have used
these measurements to calculate the expected frequencies of
hypothetical signals at harmonics and sidebands of these QPOs, as
predicted by two classes of kHz QPO models. We used a sensitive
technique to search for these harmonics and sidebands in the power
spectra of Sco X-1, but none was detected. Their power is between 10
and 100 times less than the power of the upper kHz QPO (95\,\%
confidence).

Oscillations produced close to the surface of the neutron star can be
attenuated in a scattering corona. The attenuation is generally larger
for oscillations produced by a pencil beam sweeping the surroundings
(beaming oscillations), than for oscillations due to actual changes of
the luminosity with time (luminosity oscillations; Brainerd \& Lamb
1987; Kylafis \& Klimis 1987; Kylafis \& Phinney 1989; Miller et al.
1988). The attenuation factor (the ratio of the amplitude $A_{\rm
\infty}$ of the oscillations at infinity to their original amplitude
$A_{\rm 0}$) as a function of frequency $\nu$, for a scattering corona
of a radius $R$ and optical depth $\tau$ is given by:
\begin{equation}
\label{luminosity}
\frac{A_{\rm \infty,lum}}{A_{\rm 0,lum}} = 2^{3/2} x e^{-x} + e^{-\tau}
\end{equation}
for a luminosity oscillation, and by
\begin{equation}
\label{beaming}
\frac{A_{\rm \infty,beam}}{A_{\rm 0,beam}} = \frac{2^{5/2} x e^{-x}} {1
+ \tau} + e^{-\tau},
\end{equation}
for a beaming oscillation, with $x=(3\pi \nu R \tau / c)^{1/2}$
\cite{kylafis89}.

We can use these relations to calculate the largest unattenuated
amplitudes allowed by the data in Table \ref {table_UL_obs}. Because at
each frequency the upper limit on the ratio of observed amplitudes is
$A_{\rm \infty,y}(\nu) / A_{\rm \infty,beam}(\nu_{2})$ (Table \ref
{table_UL_obs}), the upper limit on the ratio of unattenuated
amplitudes $A_{\rm 0,y}(\nu) / A_{\rm 0,beam}(\nu_{2})$, can be
obtained by minimizing 
\begin{equation}
\label{alpha}
\alpha_{\rm y}(\nu) =
\frac{ A_{\rm \infty,y}(\nu) / A_{\rm 0,y}(\nu)}
{A_{\rm \infty,beam}(\nu_{2}) / A_{\rm 0,beam}(\nu_{2})}
\end{equation}
with respect to $\tau$ and $R$, with the additional constraint that $A_{\rm
\infty,beam} (\nu_{2}) / A_{\rm 0,beam}(\nu_{2}) > 0.01$ (because the
QPO at $\nu_{2}$ is detected at an amplitude $\simmore 1$\,\%; van der
Klis et al. 1997). In eq. [\ref{alpha}], $y$ stands either for
luminosity or beaming (see eq. [\ref{luminosity}] and [\ref{beaming}]),
and we have assumed that the QPO at $\nu_{2}$ is a beaming oscillation
(Miller et al. 1998).

The upper limits to the unattenuated amplitudes of the sidebands and
harmonics of the kHz QPOs, both for luminosity and beaming
oscillations, are shown in Table \ref{table_UL_calc}. In units of the
amplitude of the QPO at $\nu_{2}$, the amplitudes of these harmonics
and sidebands range from 0.11 to 0.54 for a luminosity oscillation, and
from 0.21 to 2.16 for a beaming oscillation (assuming beaming for the
QPO at $\nu_{2}$). From the minimization of $\alpha_{\rm y}(\nu)$ at
the frequencies listed in Table \ref{table_UL_calc}, we get that
optical depths $\tau \sim 1 - 10$ and radii $R \sim 1 - 350$ km for the
putative scattering cloud in Sco X-1 would be most effective at
suppressing the sidebands and harmonics. These values are comparable to
those obtained by by Vaughan et al. \shortcite
{vaughan_lags1,vaughan_lags2}, $\tau \simless 5$ and $R \simless 80$
km, for 4U\,1608--52 from measurements of the time lags of the kHz QPOs
(see also Kaaret et al. 1999). We want to mention that the upper limits
given in Table \ref{table_UL_calc} have been calculated assuming
properties of the scattering corona that maximize the attenuation at
each frequency {\it independently}. In reality, if the corona exists it
would have fixed values of $\tau$ and $R$, so that if it is effective
at suppressing the oscillations at some frequencies, it will be less
effective at others.

In conclusion, there is a remarkable lack of harmonic and sideband
structure in the kHz QPOs of Sco X-1. None of the frequencies predicted
by the two main classes of models are detected, with upper limits
implying they are one to two orders of magnitude weaker in power than
the main peaks. This sets severe constraints on either the purity of
the kHz QPO generation mechanism in Sco X-1, or the parameters of the
scattering corona that surrounds the neutron star.

\section*{Acknowledgments}

We would like to thank various participants of the 1999 workshop on
X-ray Probes of Relativistic Astrophysics at the Aspen Center for
Physics for pleasant and extremely fruitful discussions. We are
specially grateful to Coleman Miller, Fred Lamb, Luigi Stella, and
Dimitrios Psaltis. We thank Coleman Miller, Fred Lamb and Tomaso
Belloni for comments that helped us to improve the original manuscript.
MM acknowledges comments from Mikhail Revnivtsev concerning the
dead-time effects of the PCA. This work was supported by the
Netherlands Research School for Astronomy (NOVA), the Netherlands
Organization for Scientific Research (NWO) under contract number
614-51-002 and the NWO Spinoza grant 08-0 to E. P. J. van den Heuvel.
MM is a fellow of the Consejo Nacional de Investigaciones
Cient\'{\i}ficas y T\'ecnicas de la Rep\'ublica Argentina. This
research has made use of data obtained through the High Energy
Astrophysics Science Archive Research Center Online Service, provided
by the NASA/Goddard Space Flight Center.

\bsp

\label{lastpage}


\begin{thebibliography}{}

\bibitem[\protect\citename{Bradt et al. }1993]{bradt93}
        Bradt H. V., Rothschild R. E., Swank J. H., 1993, A\&AS, 97, 355
\bibitem[\protect\citename{Brainerd \& Lamb }1987]{brainerd87}
        Brainerd J. Lamb F. K., 1987, ApJ, 317, L33
\bibitem[\protect\citename{Jahoda et al. }1996]{jahoda_pca}
        Jahoda K., Swank J. H., Giles A. B., Stark M. J., Strohmayer T.,
        Zhang W., Morgan E. H., 1996 Proc. SPIE 2808: EUV, X-ray, and
        Gamma-ray Instrumentation for Astronomy VII, p. 59. 
\bibitem[\protect\citename{Jernigan et al. }2000]{jernigan2000}
        Jernigan J. G., Klein R. I., Arons, J., 2000, ApJ, 530, 875
\bibitem[\protect\citename{Kaaret et al. }1999]{kaaret_1636}
        Kaaret P., Piraino S., Ford E. C., Santangelo A. 1999,
        ApJ, 514, L31
\bibitem[\protect\citename{Klein et al. }1996a]{klein96a}
        Klein R. I., Arons J., Jernigan J. G., Hsu J., 1996a, ApJ,
        457, L85
\bibitem[\protect\citename{Klein et al. }1996b]{klein96b}
        Klein R. I., Jernigan J. G., Arons J., Morgan E. H., Zhang W.,
        1996b, ApJ, 469, L119
\bibitem[\protect\citename{Kylafis \& Klimis }1987]{kylafis87}
        Kylafis N. D., Klimis G. S., 1987, ApJ, 323, 678
\bibitem[\protect\citename{Kylafis \& Phinney }1989]{kylafis89}
        Kylafis N. D., Phinney E. S., 1989, in Timing Neutron Stars,
        ed. H. \"Ogelman \& E. P. J. van den Heuvel, NATO ASI
        Series C262, p. 27
%\bibitem[\protect\citename{Lamb }1988]{lamb88}
%        Lamb F. K., 1988, Adv. Space Res. 8, 421
\bibitem[\protect\citename{Lamb \& Miller }1999]{lamb&miller_aspen}
        Lamb F. K., Miller M. C., 1999, paper presented at the 1999
        Aspen Summer Workshop on X-ray Probes of Relativistic
        Astrophysics
%\bibitem[\protect\citename{Leahy }1983]{leahy83}
%        Leahy D. A., Darbro W., Elsner R. F., Weisskopf M. C.,
%        Sutherland P. G., Kahn S., Grindlay J. E., 1983, ApJ,
%        266, 160
\bibitem[\protect\citename{M\'endez et al. }1998a]{mendez1608_1}
        M\'endez M., et al., 1998a, ApJ, 494, L65
\bibitem[\protect\citename{M\'endez et al. }1998b]{mendez1608_2}
        M\'endez M., van der Klis M., Wijnands R., Ford E. C.,
        van Paradijs J., Vaughan B. A., 1998b, ApJ, 505, L23
%\bibitem[\protect\citename{M\'endez et al. }1999]{mendez_1608_3}
%        M\'endez M., van der Klis M., Ford E. C., Wijnands R.,
%        van Paradijs J., 1999, ApJ, 511, L49
\bibitem[\protect\citename{Miller }2000]{miller_bologna}
        Miller M. C., 2000, in Proc. of the Bologna Conf. ``X-ray
        Astronomy 1999: Stellar Endpoints, AGN, and the Diffuse X-ray
        Background'', in press
\bibitem[\protect\citename{Miller et al. }1998]{miller_sonic}
        Miller M. C., Lamb F. K., Psaltis D., 1998, ApJ, 791
\bibitem[\protect\citename{Osherovich \& Titarchuk }1999]{osherovich99}
        Osherovich V., Titarchuk L., 1999, ApJ, 522, L113
\bibitem[\protect\citename{Psaltis }1999]{psaltis_aspen}
        Psaltis D., 1999, paper presented at the 1999 Aspen Summer
	Workshop on X-ray Probes of Relativistic Astrophysics
\bibitem[\protect\citename{Psaltis \& Norman }2000]{psaltis_norman}
        Psaltis D., Norman C., 2000, ApJ, in press
\bibitem[\protect\citename{Stella }1999]{stella_aspen}
        Stella L., 1999, paper presented at the 1999 Aspen Summer 
        Workshop on X-ray Probes of Relativistic Astrophysics
\bibitem[\protect\citename{Stella \& Vietri }1999]{stella99}
        Stella L., Vietri M., 1999, PRL, 82, 17
\bibitem[\protect\citename{Strohmayer, Zhang \& Swank }1996]{stroh_iauc}
        Strohmayer T., Zhang W., Swank J., 1996, IAU Circ. 6320
\bibitem[\protect\citename{van der Klis }1995]{vanderklis95}
        van der Klis M., 1995, in X-ray Binaries, eds. W.H.G. Lewin, J.
	van Paradijs \& E.P.J. van den Heuvel
        (Cambridge: Cambridge Univ. Press) p. 252
\bibitem[\protect\citename{van der Klis }2000]{vdk_araa}
        van der Klis M., 2000, ARA\&A in press (astro-ph/0001167)
\bibitem[\protect\citename{van der Klis et al. }1996a]{vdk_iauc}
        van der Klis M., Swank J., Zhang W., Jahoda K., Morgan E.,
        Lewin W., Vaughan B., van Paradijs J., 1996a, IAU Circ. 6319
\bibitem[\protect\citename{van der Klis et al. }1996b]{vdk_sco_1}
        van der Klis M., Swank J. H., Zhang W., Jahoda K., Morgan E. H.,
        Lewin W. H. G., Vaughan B., van Paradijs J., 1996b, ApJ, 469, L1
\bibitem[\protect\citename{van der Klis et al. }1997]{vdk_sco_2}
        van der Klis M., Wijnands R. A. D., Horne K., Chen W., 1997,
	ApJ, 481, L97
\bibitem[\protect\citename{Vaughan et al. }1997]{vaughan_lags1}
        Vaughan B. A., et al., 1997, ApJ, 483, L115
\bibitem[\protect\citename{Vaughan et al. }1998]{vaughan_lags2}
        Vaughan B. A., et al., 1998, ApJ, 509, L145
\bibitem[\protect\citename{Zhang et al. }1995]{zhang_dead-time}
        Zhang W., Jahoda K., Swank J.H., Morgan E. H., Giles, A. B.,
        1995, ApJ, 449, 930

\end{thebibliography}
\end{document}